# Light transport and vortex-supported wave-guiding in micro-structured optical fibres

ANDREY PRYAMIKOV[1], GRIGORY ALAGASHEV[2], GREGORY FALKOVICH[3,5], AND SERGEI TURITSYN[4,5,*]

[1]*Fibre Optics Research Center of Russian Academy of Sciences, Moscow, Russia*
[2]*D. Mendeleev University of Chemical Technology of Russia, Moscow, Russia*
[3]*Weizmann Institute of Science, Rehovot 76100, Israel*
[4]*Aston Institute of Photonic Technologies, Aston University, Birmingham, UK*
[5]*Aston-Novosibirsk Centre for Photonics, Novosibirsk State University, Novosibirsk, Russia*
*\* s.k.turitsyn@aston.ac.uk*

**Abstract:** In hydrodynamics, vortex generation upon the transition from smooth laminar flows to turbulence is generally accompanied by increased dissipation. However, plane vortices can provide transport barriers and decrease losses, as it happens in numerous geophysical, astrophysical flows and in tokamaks. Photon interactions with matter can affectlight transportin waveguides in unexpected and somewhat counterintuitive ways resembling fluid dynamics. Here, we demonstrate dramatic impact of light vortex formation in micro-structured optical fibres on the energy dissipation. We show possibility of vortices formation in both solid core and hollow core fibres on the zero flow lines in the cladding. We find that vortices reduce light leakage by three orders of magnitude, effectively improving wave guiding. A strong light localization based on the same principle can also be achieved in the negative curvature hollow core fibres.

## 1. Introduction

Transverse confinement of a flow is fundamental to many fields of science and technology. Decreasing momentum losses through the pipe walls reduces drag and can save us trillions in the energy cost of oil and gas transport. Decreasingheat losses through tokamak walls is crucial forthermonuclear fusion. Wave guiding of photons in optical fibre is important for optical communications (and Internet), high power lasers, beam delivery and other optical technologies. In standard step-index fibre transversal confinement is ensured by the total internal reflection, but in micro-structured fibres decreasing light leakage is a challenge.

Appearance of vortices in a flow can have opposite effects on confinement. In pipe and channel flows, from industrial to cardiovascular systems, generation of vortices generally increases dissipation while vortex suppression can lead to drag reduction. On the contrary, in rotating, magnetized systems and in fluid layers, quasi-two-dimensional vortices have separatrices, serving as transport barriers. Most dramatic example is the transition from low to high confinement in tokamak when zonal vortex flow suppresses heat transfer to the walls. Here we show how generation of optical vortices can dramatically improve light confinement in micro-structured fibres.

Though optics and hydrodynamics look distinct, wave dynamics and underlying mathematical models have many qualitative and quantitative similarities [1-5]. In particular, in nonlinear fibre optics a number of physical effects have been observed, that closely resemble nonlinear hydrodynamic problems, including modulations instability, oceanic rogue waves, localised nonlinear structures, shock waves, optical turbulence [1-12]. These similarities provide possibilities to transfer knowledge and techniques and observe beautiful and non-evident connections between the two diverse fields. The present work reveals one such connection through the analysis of formation of vortex structures in optical fibres and its impact on the energy flows.

For quantum and wave phenomena described by a complex field, vortex is an amplitude-zero topological phase defect [1-12]. For instance, an anomalous light transmission through a subwavelength slit in a thin metal plate is accompanied by wave-guiding and phase singularities – vortices of the optical power flow [13]. In nonlinear media, the vortices can exist in the form of topological solitons [14].

Three main features associated with OV are zero intensity and phase indefiniteness in the center (pivot point), and a screw dislocation of the wave front [15, 16]. In two dimensions, phase singularities occur when three or more plane waves or two Gaussian pulses interfere and light vanishes at some points [17]. Phase rotates by $2\pi$ around the zero-intensity point, which leads to a circulation of the optical energy. A circular flow of energy leads to the ability of optical vortices to carry angular momentum (AM).

The classical AM is well studied or the monochromatic waves in free space. In recent years, AM transfer in dielectrics and fibres has attracted much interest [18-23]. To date, the OVs in waveguides have been considered mostly for eigenmodes of step-index or graded - index fibres [24 -26]. In this work, we make connection between OVs formation and their impact on the fibre loss, which was not studied before.

## 2. Results

We examine the linear OVs that occur in the cladding of micro-structured optical fibres (MOFs, all solid photonic band gap fibres (ASBGF) [27] and a new type of hollow core fibres - negative curvature hollow core fibres (NCHCF) [28-30]. OVs considered here arise in the transverse component of the Poynting vector of the core modes, which determines the losses of these modes in the micro-structured fibres. The transverse component of the Poynting vector always has an uncertainty in direction and zero value at the origin. Rotational symmetry of the cladding elements arrangement in the micro-structured fibres defines the azimuthal energy fluxes of the core modes. This energy flux leads to the formation of additional vortices both inside the cladding elements and in the space between them (Fig. 1). It is well known that phase dislocations in the electromagnetic wave structure in free space occur when the real and imaginary parts of the field strength are simultaneously equal to zero [31]. It was shown in [32] that the singular points of the stream lines of the transverse Poynting vector $\vec{P}_{transv}$ of the core modes in a cross section of the graded – index fibre are found from the equations $P_x(x,y) = P_y(x,y) = 0$, where $P_x$ and $P_y$ are the components of $\vec{P}_{transv}$. We will demonstrate that these conditions also define the positions of the OV centers in the cladding of ASBGFs and NCHCFs. Effectively, we demonstrate OV induced wave-guiding effect, in which photons passing through the optical fibre excite the circulating power currents that impact light transport decreasing the overall losses. The formation of the OV in the cladding capillary walls of NCHCFs leads to a strong light localization in the hollow core, which makes it possible to transmit radiation in the mid IR spectral range [33].

## 3. Vortex – supported wave - guiding in all solid band gap fibres

Dislocations of monochromatic waves are stationary in space and form isolated interference fringes. The phase of the field is undefined on the zero amplitude lines and changes by $\pi$ when crossing them. The wave front dislocations of coherent radiation can be characterized by zero field amplitude in its center or along the dark rings where the field amplitude vanishes.The origin of the dark rings and dislocation centers is boundary diffraction and destructive interference.

For such micro – structured optical fibres as ASBGFs or NCHCFs the electric and magnetic fields of the core modes $\vec{E}(r,\phi)e^{i(\beta z-\omega t)}$ and $\vec{H}(r,\phi)e^{i(\beta z-\omega t)}$ can be described by specifying the axial components of the fields $E_z$ and $H_z$, where $\omega$ denotes the angular frequency and $\beta$ is propagation constant of the core mode. Then, the wave equation for each axial component is the Helmholtz equation:

$$\vec{\nabla}^2 F + k_t F = 0,$$

where $F = E_z$ or $F = H_z$ and $k_t = \sqrt{k_0^2 n_i^2 - \beta^2}$ is the transverse component of the wavevector, $k_0 = 2\pi/\lambda$ and $n_i$ is a refractive index of the cladding element or surrounding area. In the case of micro - structured fibres it is possible to separate the "longitudinal" phase of the core mode $e^{i\beta z}$ and the "transversal" phase.

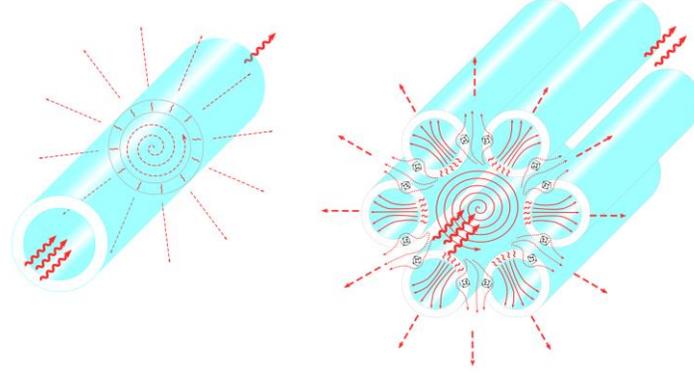

Fig.1. Vortex – supported guidance in micro – structured fibres. (left) Monochromatic radiationin the leaky core mode of the hollow capillary. The axial component of the Poynting vector is from left to right (wavy red arrows). The losses of the mode are determined by the transverse component of the Poynting vector whosestreamlinesare represented by dashed red arrows. The lossesare high because the streamlines of the transverse component are refracted the hollow core boundary and further in the empty space are directed along the radial direction. (right) Hollow core micro - structured fibre. The streamlines of the transverse component of the Poynting vector are deviatedby the optical vortices in the cladding, where the transverse component of the Poynting vector turns into zero, which leads to loss reduction.

Each component of the core mode fields in the vicinity of the cladding element is a linear superposition of an infinite set of cylindrical harmonics. Axial components of the fields can be expressed as [34]:

$$E_z = \sum_{m=-\infty}^{+\infty} A_m F_m(k_t r) e^{im\phi}, \qquad (1)$$

$$H_z = \sum_{m=-\infty}^{+\infty} B_m F_m(k_t r) e^{im\phi},$$

where $F_m(r)$ is Bessel function of $1^{th}$ order if $r < a$ and Hankel function of $1^{th}$ order if $r > a$, $a$ is the radius of the cladding dielectric cylinder. For Hankel function of $1^{th}$ order the radiation condition at infinity is satisfied.

All transversal components of the core mode fields can be expressed in terms of the axial component using well known relations [35]. For example, azimuthal components of the core mode fields are:

$$E_\phi = \frac{i}{k_t^2}\left(\frac{\beta}{r}\frac{\partial E_z}{\partial \phi} - k_0 \frac{\partial H_z}{\partial r}\right),$$ (2)

$$H_\phi = \frac{i}{k_t^2}\left(\frac{\beta}{r}\frac{\partial H_z}{\partial \phi} + k_0 n_i^2 \frac{\partial E_z}{\partial r}\right).$$

All the details of calculation of the expansion coefficients of the cylindrical harmonics in (1) and the complex propagation constants of the core modes performed using the Multipole method can be found in [34] and in the section Methods. In addition, we performed the same calculations using COMSOL software package.

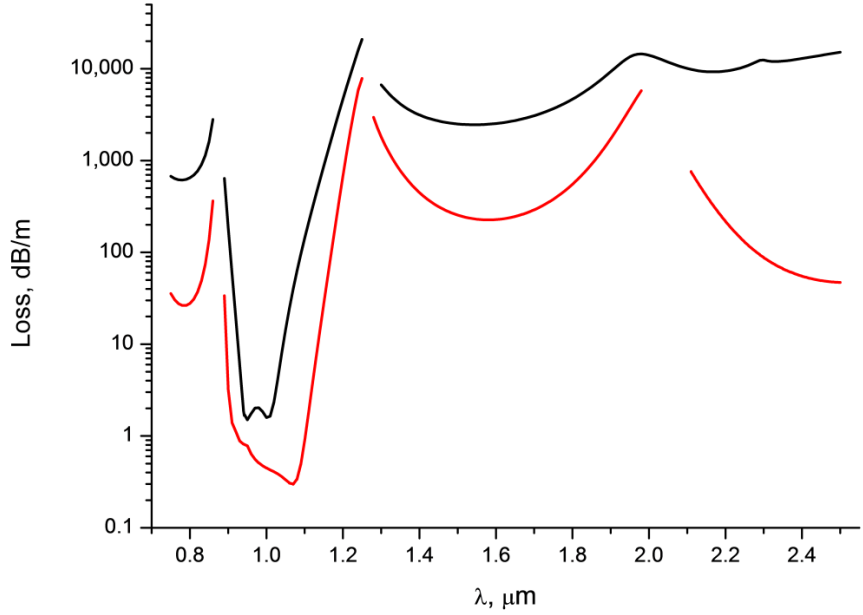

Fig. 2. Loss dependence of the fundamental core mode on the wavelength for all solid band gap fibres with one (black) and two rows (red) of the cladding rods. The geometrical and material parameters of the fibres are in the text.

We considered all solid band gap fibres with one and two rows of the cladding dielectric rods with a refractive index contrast $\Delta n = 0.05$ between refractive index of the cladding rods and the surrounding glass matrix ($n_{glass} = 1.45$). The arrangement of the cladding rods has a hexagonal structure. The value of a pitch of $\Lambda = 12$ μm (distance between the centers of the cladding rods), and the ratio of $d/\Lambda = 0.33$, where $d$ is the cladding rod diameter.

Let us consider the light leakage from thefibre core and calculate loss dependencies on the wavelengthof the fundamental core mode in several transmission bands. The loss dependencies on the wavelength for both ASBGFs are shown in Fig. 2. It is seen that there are relatively narrow transmission bands in which the losses are three orders of magnitude lower than in the rest of the spectrum.

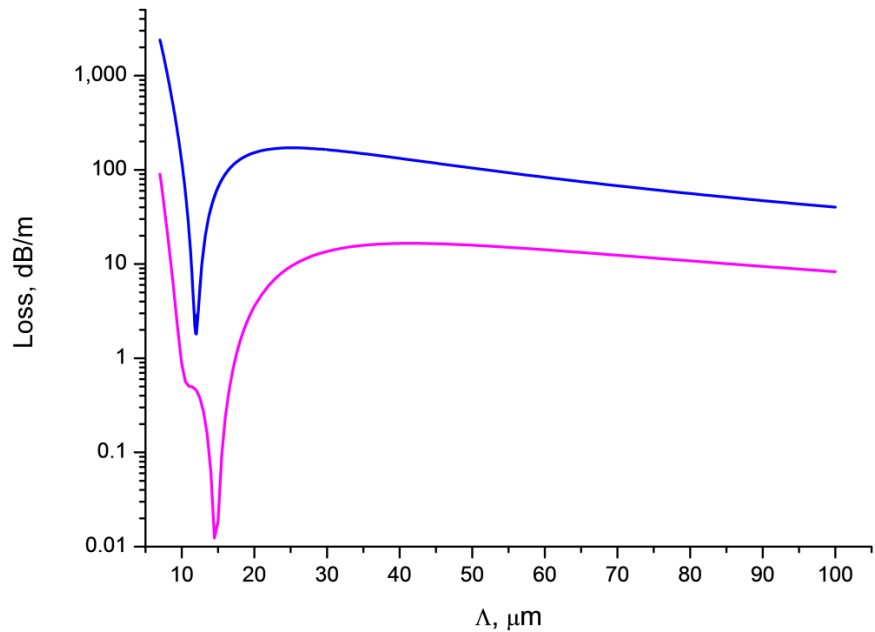

Fig. 3. The loss dependencies of the fundamental core mode on the value of the pitch for all solid band gap fibres described in the text. The blue curve for one row, the magenta curve for two rows of the cladding rods.

This is especially true for the fibres with one row of the cladding rods. The same resonant loss reduction can be observed if we fix the wavelength, for example, in the minimum of losses at a wavelength of $\lambda = 1$ μm, and change the value of the pitch $\Lambda$. The calculation results are shown in Fig. 3. As in Fig. 2, there is a sharp decrease in losses by several orders of magnitude in both cases in a narrow range of values of the pitch. It is clear that this substantial decrease in losses for both fibres can be only associated with the narrow spectral regions and with specific values of the pitch.

Light leakage of the core modes can be characterized by the distribution of projection of the transverse component of the Poynting vector on the radius - vector drawn from the origin $P_r$. The values of the projection of the transverse component of the Poynting vector for the fundamental core mode were calculated at the wavelengths of $\lambda = 1$ μm, 1.5 μm and a value of pitch of $\Lambda = 12$ μm.

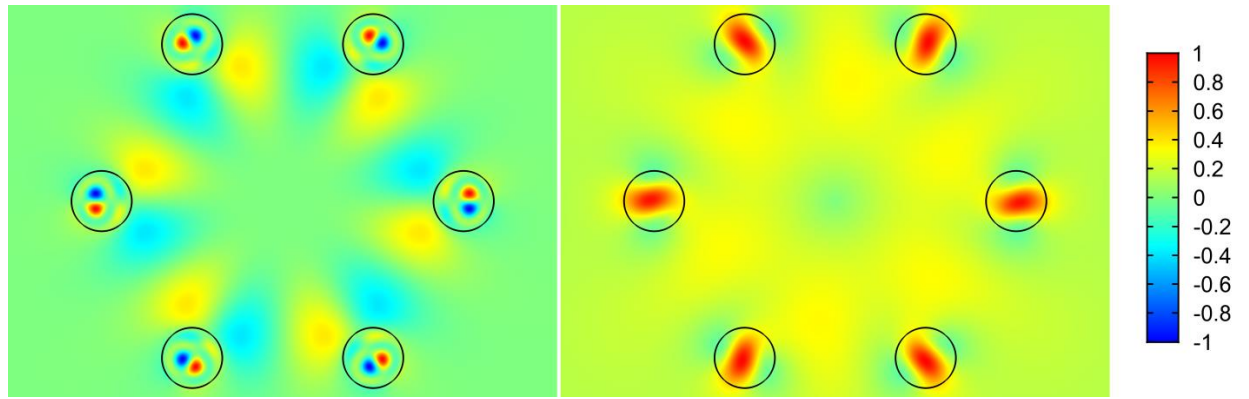

Fig. 4. The map of the radial component of the Poynting vector for all solid band gap fibres with one row of rods. (left)λ = 1 μm and Λ = 12 μm (see Fig. 1); (right) λ = 1.5 μm.

As in the case of polygonal waveguides [36], the distribution of projection of the transverse component of the Poynting vector on radius – vector has periodic alternating characterwhich points to a vortex structure of the core mode fields in the cladding (Fig. 4). The vortex structures of the core mode fieldsfor bothdistributions of the radial projection of the transverse component of the Poynting vector are different at different wavelengths (Fig. 4).In the case of minimal losses (Fig. 4(left)), the OVs centers are located inside the cladding rods, in the case of large losses in the fibre (Fig. 4(right)), the OVs are located at the boundaries of the cladding rods.Thus, the leaky radiation of the fundamental core modemovesalong different trajectories in the cladding. It is well established that the single fundamental property of the optical vortex formation is the rotation of the Poynting vector (energy rotation) around the phase dislocation (the OV core) [15, 16]. To clearly demonstrate the OV formation and the formation of phase dislocations of various structures in the cladding of the fibre with one ring of the cladding rods (Fig. 4) let us consider the distribution of the stream lines of the transverse Poynting vector component of the fundamental core modes $\vec{P}_{transv}$ and the lines of zero components of $\vec{P}_{transv}(x, y) = \{P_x(x, y) = 0; P_y(x, y) = 0\}$ at wavelengths of $\lambda = 1$ µm and $\lambda = 1.5$ µm (Fig. 2). The distribution for the wavelength of $\lambda = 1$ µm is shown in Fig. 5.

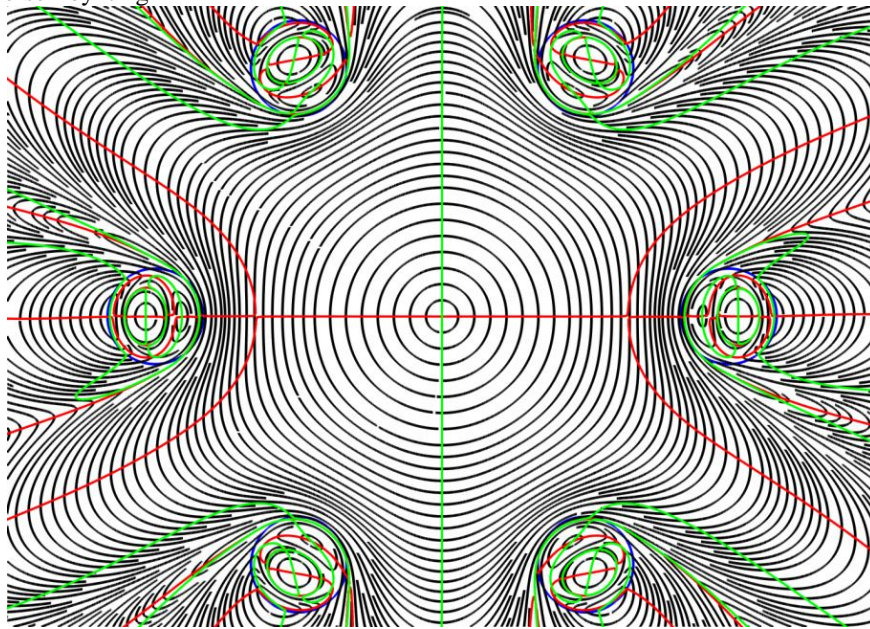

Fig. 5. The streamlines of the transverse component of the Poynting vector (grey lines) and the lines of zero intensity $\vec{P}_{transv} = \{P_x = 0; P_y = 0\}$ ($P_x = 0$ (red lines), $P_y = 0$ (green lines)) for all solid band gap fibre shown in Fig. 3 at a wavelength of $\lambda = 1$ µm.

The structure of the Poynting vector streamlines shown in Fig. 5 points to the formation of the dislocation lines around which the leaky radiation of the fundamental core mode rotates. The vortex centers are formed at the intersection points of the curves $P_x(x, y) = 0$ and $P_y(x, y) = 0$. Moreover, it can be seen from Fig. 5 that there are closed areas, where light rotates around the dislocation lines. In addition, there are areas of the cladding where the two curves are very close to each other and form the ring phase dislocations inside the rods [38].

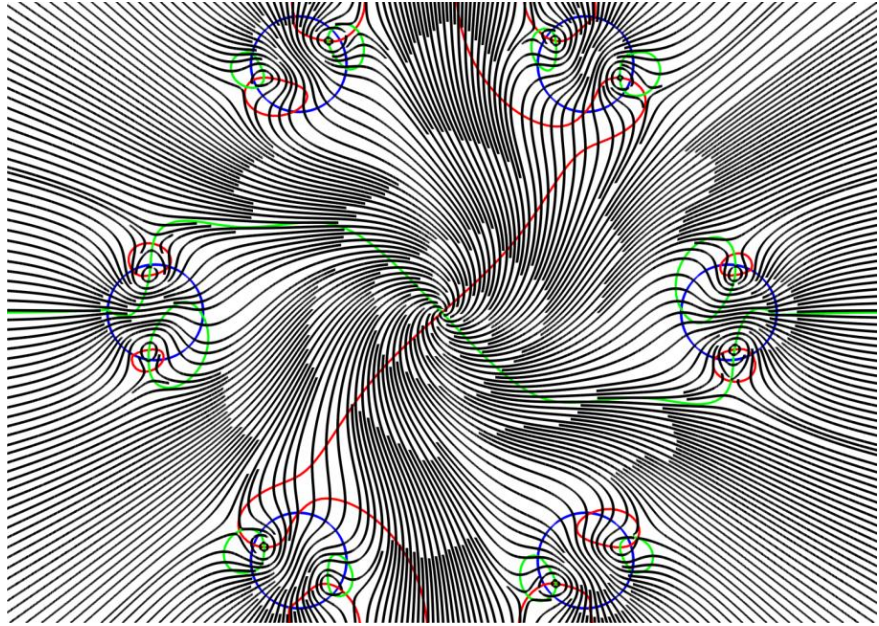

Fig. 6. The streamlines of the transverse component of the Poynting vector (grey lines) and the zero lines $\vec{P}_{transv} = \{P_x = 0; P_y = 0\}$ ($P_x = 0$ (red lines), $P_y = 0$ (green lines)) for all solid band gap fibre shown in Fig. 3 at a wavelength of $\lambda = 1.5$ µm.

Although the streamlines distribution of the transverse component of the Poynting vector at a wavelength of $\lambda = 1.5$ µm forms the OVs (Fig. 6), the energy of the core mode efficiently flows through different paths in the cladding and rotates only in small regions near the OV centers at the rod boundaries. Moreover, a major part of the core mode radiation flows through the cladding rods (Fig. 4) which, in contrast to the previous case, cannot serve as effective reflectors for the leaky radiation. This leads to large losses in the fibre (Fig. 2 and Fig. 3).

Let us now consider the phase distribution of the transverse component of core mode electric field, for example, $E_x$. According to the general principles of the optical vortices theory [15, 16], the phase distribution of the transverse component of the core mode electric field should experience a jump in π when passing through the boundary of the closed area around the dislocation lines. We calculated the phase distribution of $E_x$ in the cladding of both fibres for two wavelengths of $\lambda = 1$ µm and $\lambda = 1.5$ µm (Fig. 2).

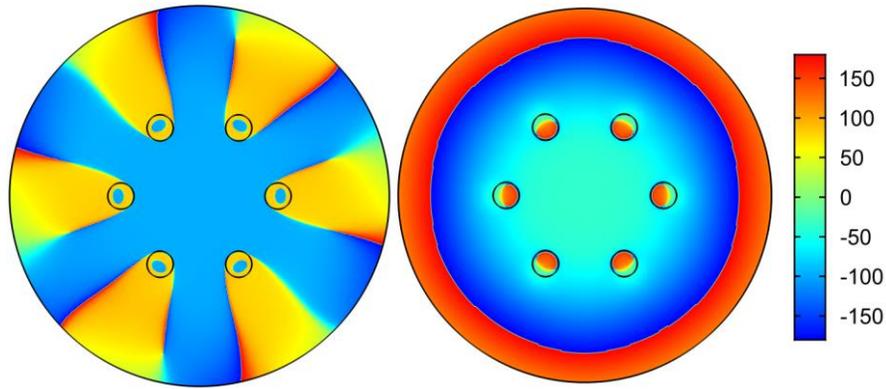

Fig. 7. (left) the phase distribution of the transverse electric field component $E_x$ for the fundamental core mode of a fibre with one row of the cladding rods at a wavelength of $\lambda = 1$ µm; (right) the phase distribution of the transverse electric field component $E_x$ for the fundamental core mode of the fibre with one row of the cladding rods at a wavelength of $\lambda = 1.5$ µm.

The phase distribution of $E_x$ for the fibre with one row of the cladding rods is shown in Fig. 5. The vertical scale represents the phase values in degrees. It can be seen from Fig. 7(left) that the OVs located in the cladding rods at wavelength of $\lambda = 1$ µm have a corresponding phase jump $\pi$ for the distribution of $E_x$. The phase distribution of $E_x$ at a wavelength of $\lambda = 1.5$ µm has a qualitatively different character Fig. 7(right). In this case, the phase distribution does not experience a pronounced jump in $\pi$ in the cladding rods, so the ring phase dislocations for the fundamental core mode don't form in the cladding rods. This difference in the core mode field structure leads to a difference in the loss level at these two wavelengths. It is possible to demonstrate that in the case of ASBGFs with two rows of the cladding rods (Fig. 2 - 4) the loss reduction is also determined by the phase ring dislocations in the rods.

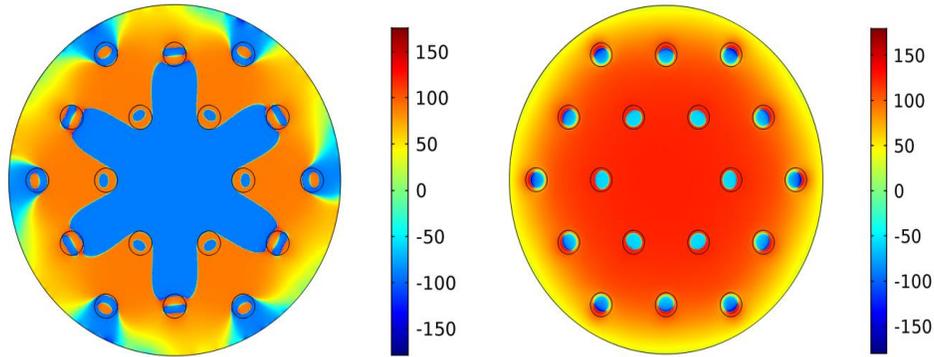

Fig. 8. (left) the phase distribution of the transverse electric field component $E_x$ for the fundamental core mode of a fibre with two rows of the cladding rods at a wavelength of $\lambda = 1$ µm; (right) the phase distribution of the transverse electric field component $E_x$ for the fundamental core mode of the fibre with two rows of the cladding rods at a wavelength of $\lambda = 1.5$ µm.

In this case, the phase distribution of the transverse component of the core mode electric field experience a pronounced jump in $\pi$ when passing through the boundary of the closed area around the dislocation lines in both rows of the cladding rods at wavelength of $\lambda = 1$ µm (Fig.

8 (left)). The ring phase dislocations don't form in the case of big loss at wavelength of λ = 1.5 µm (Fig. 8 (right)).

Figures 5 - 6 demonstrate qualitative (topological) difference between two cases. Only isolated vortices exist for λ = 1.5, which makes the energy to spiral out without returning. On the contrary, the green and red lines in Fig.5 cross not only at isolated points but also coincide along the whole radial lines, which are thus vortex lines having phase jump in the left panel of Fig.7 and Fig.8. Those vortex lines provide for energy recirculation and improved confinement for λ = 1 µm.

## 4. Vortex – supported wave - guiding in the negative curvature hollow core fibres

To demonstrate the vortex - supported wave – guidingin the NCHCFs and its distinction from the wave – guiding in waveguides withcontinuous rotational symmetry ofthe core – cladding boundary, let us consider the loss dependence of the fundamental air core mode on the wavelength for three waveguide micro - structures.

The first one is a single capillary with a wall thickness of 0.65 µm and the air core diameter of 14 µm. The refractive index of the capillary wall is equal to 1.5 as in the case of ASBGF considered in Section 2.

The second one is a waveguide consisting of two capillaries nested in one another and having a common center.The internal capillary has the same parameters as the single capillary described above.The outer capillary has a diameter of 18.8 µm and the same wall thickness as the internal capillary. The light localization in both waveguides can be described within the ARROW model [39]. The capillary wall can be considered as Fabry – Perrot resonator which either passes radiation outside (condition of the resonant regime is $k_t d = \pi m$, where $k_t$ is the transverse component of the air core mode wavevector, $d$ is the capillary wall thickness and $m$ is an integer) or reflects it into the air core (condition of the anti – resonant regime is $k_t d = \pi(m+1/2)$). In this case, the losses are consequently reduced with the addition of each new cladding layer (Fig. 9). The NCHCF has 6 cladding capillaries with a wall thickness of 0.65 µm, the inner diameter of 8.3 µm and the air core diameter is 14 µm. The distance between the nearest points of the adjacent cladding capillaries is 3 µm.

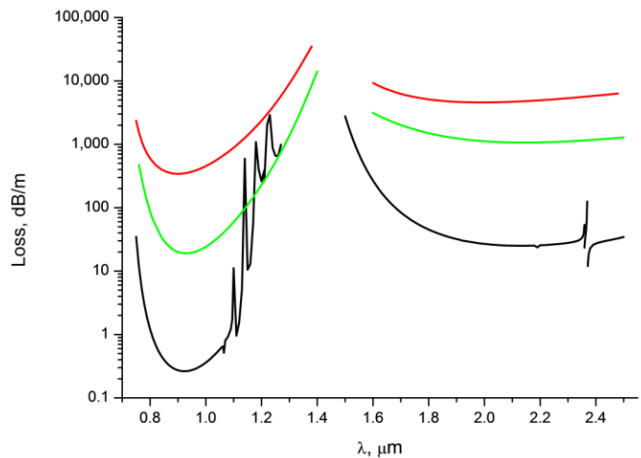

Fig.9. Loss dependence of the fundamental air core mode on the wavelength for three waveguide microstructures: NCHCF with 6 capillaries in the cladding (black), a single capillary waveguide (red) and a double capillary waveguide (green).

The losses of NCHCF with 6 cladding capillaries are approximately two orders of magnitude lower than in the double capillary fibre at the point of minimum loss (Fig. 9). This difference in losses can be explained by the difference in the leakage process for the fundamental air core mode in both cases. Another dissimilarity pertains to the loss behavior of the NCHCF at the long wavelength edge of the short wavelength transmission band (Fig. 9) and originates from the coupling between the air core and the cladding modes [40].

As in the case of ASBGFs, the leakage process for the fundamental air core mode can be characterized by the streamlines of the transverse component of the Poynting vector and the lines of zero components of $\vec{P}_{transv} = \{P_x(x, y) = 0; P_y(x, y) = 0\}$. The corresponding distribution of the streamlines of $\vec{P}_{transv}$ and the points of intersection of $P_x(x, y) = 0$ and $P_y(x, y) = 0$ are shown in Fig. 10 for the internal capillary of the double capillary waveguide and the single cladding capillary of the NCHCF whose losses are shown in Fig. 9. The calculations were carried out at the center of the second transmission band at a wavelength of $\lambda = 0.9$ µm (Fig. 9). It can be seen from Fig. 10 (left) that in the case of the double capillary waveguide the curves $P_x(x, y) = 0$ and $P_y(x, y) = 0$ have no intersection points.

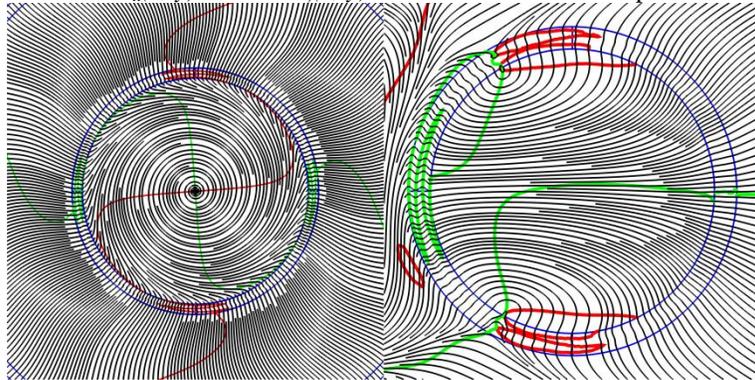

Fig. 10. Streamlines of the transverse component of the Poynting vector (grey) of the fundamental core mode and its zerolines ($P_x = 0$ (red lines), $P_y = 0$ (green lines)) for the internal capillary of the double capillary fibre (left) and (right) for the single cladding capillary of NCHCF with 6 cladding capillaries at a wavelength of $\lambda = 0.9$ µm (Fig. 9).

Thus, the OV formation in the capillary wall is impossible. It was shown in [36] that in this case the radial projection of the transverse component of the Poynting vector $P_r$ must be positive along the whole core – cladding boundary of the capillary. For the double capillary fibre this distribution is shown in Fig. 11(left) for the sum of two orthogonally polarized fundamental air core modes. In all points of the cladding $P_r > 0$ and streamlines are directed only along the radial direction. The distribution of $P_r$ of the sum of two orthogonally polarized fundamental air core modes on the azimuthal angle $\varphi$ along the circle with a radius of $R = 10.6$ µm is shown in Fig. 11(right). The magnitude of $P_r$ is normalized to the value of the axial component of the Poynting vector for the sum of two orthogonally polarized fundamental air core modes, which is taken to be one watt. The distribution of $P_r$ has no features on the azimuthal angle (Fig. 11(right)).

A different situation is observed in the case of the NCHCFs. The OVs centers are formed at the intersection points of the curves $P_x(x, y) = 0$ and $P_y(x, y) = 0$, where the distance between the cladding capillaries is minimal (Fig. 10(right)). They occur both at the boundary and inside the capillary wall, and the flowing energy of the fundamental air core mode rotates around them. As in the case of the ASBGFs (Fig. 4), part of the core mode energy flux must changesits direction in the regions between the cladding capillaries and then returns to the air core of NCHCF. To confirm this assumption let us consider a distribution of $P_r$ at a wavelength of 0.9 μm is shown in Fig. 12(left).

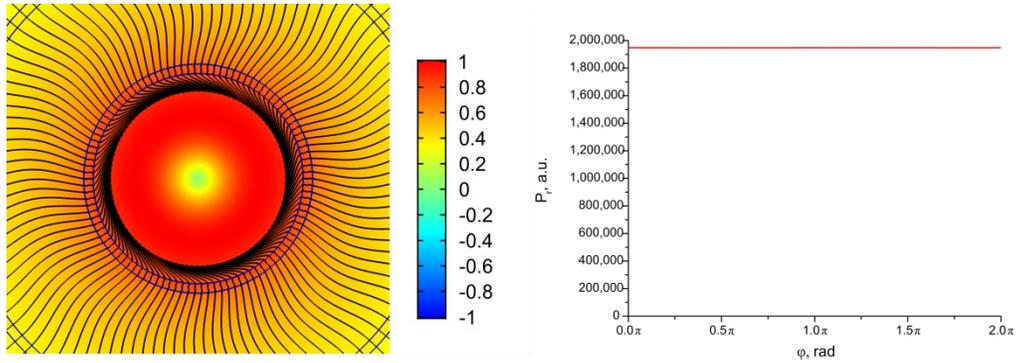

Fig. 11. (left) the distribution of projection of the transverse component of the Poynting vector on the radius – vector drawn from the origin for the double capillary fibre at a wavelength of $\lambda = 0.9$ μm (Fig. 9); (right) the distribution of this projection on azimuthal angle φalong the circle with a radius of $R = 10.6$ μm.

It can be seen from Fig. 12(left) that due to the presence of the OVs in the cladding capillary walls there areregions between the capillaries, where $P_r = 0$ or $P_r < 0$. Due to the presence of the OVs, the flowing energy of the fundamental air core mode passes only through limited segments of the cladding capillary wall surfaces located closest to the center of the core and only a small part of this energy passes through the space between the cladding capillaries. To confirm this conclusion let us consider the dependence of $P_r$ on the azimuthal angle $\varphi$ along a circle with a radius of 10.6 μm passing near the centers of the OVs (Fig. 12(right)). It can be seen from Fig. 12(right) that the distribution and value of $P_r$ are largely determined by the OV locations in the cladding. Radiation leakage from the NCHCF also occurs in the anti – resonant regime at a wavelength of $\lambda = 0.9$ μm and the level of $P_r$ in this case is much lower than in the case of the double capillary fibre (Fig. 11(right)). In this case, the normalization of $P_r$ is performed as in the case of the double capillary fibre. In [41] this regime of light localization was called the local ARROW mechanism (only a limited part of the capillary wall reflects light in the anti – resonant regime).

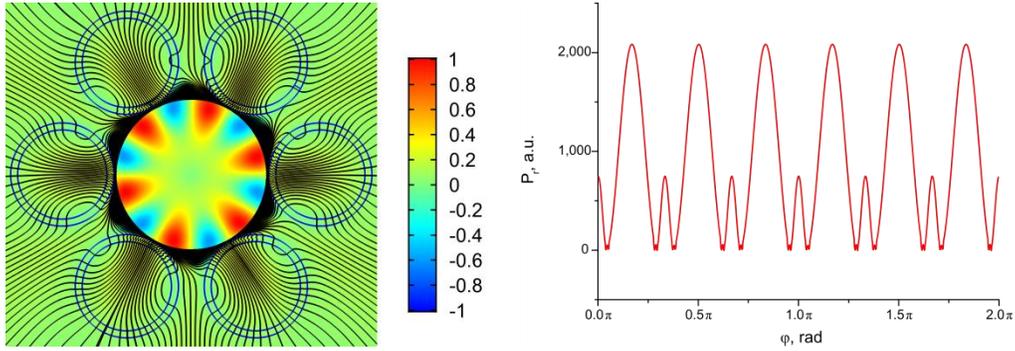

Fig. 12. (left) the distribution of projection of the transverse component of the Poynting vector on the radius – vector drawn from the origin for NCHCF with 6 cladding capillaries and streamlines of the transverse component of the Poynting vector (black lines) at a wavelength of λ = 0.9 μm (Fig. 9); (right) the distribution of this projection on azimuthal angle φ along the circle with a radius of R = 10.6 μm.

In recent paper [42] using technique of transverse power flow streamlines visualization for the air core modes of the negative curvature hollow core fibres proposed in [37] it was demonstrated a possibility of reducing losses with only a modest increase in fabrication complexity. When approaching the resonant condition for the cladding elements, for example, at a wavelength of λ = 0.75 μm (Fig. 9), the cladding capillary walls become more transparent for the outgoing radiation and the total loss level increases (Fig. 13). Due to the presence of the OVs and the reflective properties of the cladding capillaries this increase in losses is not significant compared to the increase in losses for the double capillary waveguide (Fig. 9). In this case, the vortex structure of $\vec{P}_{transv}$ is preserved which leads to an inhomogeneous distribution of the outgoing energy flux depending on the azimuthal angle and reduces the leakage loss growth (Fig. 13(right)).

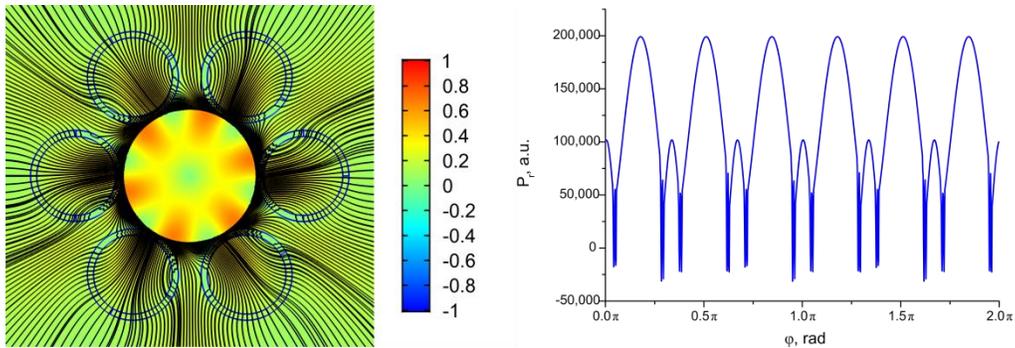

Fig. 13. (left) the distribution of projection of the transverse component of the Poynting vector on the radius – vector drawn from the origin for NCHCF with 6 cladding capillaries and streamlines of the transverse component of the Poynting vector (black lines) at a wavelength of λ = 0.75 μm (Fig. 9); (right) the distribution of this projection on azimuthal angle φ along the circle with a radius of R = 10.6 μm.

The geometrical parameters of both NCHCFs and ASBGFs have a strong impact on the OV formation in the cladding and, consequently, on wave – guide properties of the micro – structured fibres. By changing the distance between the walls of adjacent cladding capillaries, the number of the capillaries in the cladding or the distance between their centers, it is possible to control the positions of the vortices in the cladding and the level of light localization in the air core [36]. Moreover, as it was shown in [37], the introduction of the supporting tube does not significantly change thevortex formation mechanism in NCHCFs.

## 5. Conclusion

We have studied impact of transversal light vortex formation in micro-structured optical fibres on light energy leakage. It is demonstrated thatcreation of vortices inside the zero intensity lines can reduce losses by three orders of magnitude. Appearance of vortices suppresses light leakage effectively improving wave-guiding. Depending on the type of rotational symmetry of the arrangement of the cladding elements, minimal losses are achieved when centers of the phase dislocations are located in the cladding element wall. The same underlying physical mechanismprovides for a strong light localization in the negative curvature hollow core fibres. The key point of this general concept is that the right arrangement of vortices in the fibre cladding produces a balance between the outward and inward energy flux of thecore mode.We anticipate that this mechanism can be used to develop advanced low- loss fibres across spectral range for various applications, from telecom to high power lasers and optical beam delivery.

**Funding**

The work of SKT was supported by the Russian Science Foundation (Grant No. 17-72-30006). The work of AP and GA was supported by the Russian Science Foundation (Grant No. 18- 19 - 00733).

The authors declare that there are no conflicts of interest related to this article.

**Methods**

To describe vortex formation in the transverse component of the Poynting vector of the leaky core modes of micro-structured optical fibres, we employed two computational methods accounting for all electric and magnetic components of the core mode fields and their complex propagation constants. First, we use a multipole expansion method to perform full-vector modal calculations of the micro-structured optical fibres with circular rods or capillaries in the cladding. This is an efficient approach for micro-structured fibres with circular cladding elements. For a given time dependence is $exp(-i\omega t)$, the core mode fieldsare expressed via harmonics.In the neighborhood of the circular cladding element the axial components are presented using local cylindrical coordinates ($r_i$, $\varphi_i$), where $i$ is a number of the circular cladding element. In the case of the cladding rods, the axial field components can expressed in terms of Bessel functions of the first kind ($J_m$) inside the rod. In the case of the cladding capillary the axial components present the sum of Hankel functions of the first and second kind ($H_m^{(1)}$ and $H_m^{(2)}$) in the capillary wall and Bessel functions of the first kind ($J_m$) in the capillary hollow core. Matching the boundary conditions for radial and azimuthal components of the core mode fields in two domains leads to a matrix equation. It should be noted that the source-free $J_m$ parts of the expansion in the neighborhood of the cladding rod or the capillary i are due to $H_m^{(1)}$ fields radiated from the cladding rods or capillaries $j \neq i$, and in order to obtain the matrix equation when applying the boundary conditions it is necessary to use Graf's addition theorem. The determinant of the matrix defines the propagation constants of the core modes $\beta$, and the associated null vectors determine the modal fields. We cross-check the results of our calculations of the propagation constants of the leaky core modes using a commercial software package COMSOL Multiphysics based on finite element method (FEM). The complex propagation constants and distribution of the fields of the leaky core modes have been found by solving the eigenvalue problem for the wave equation for the two types of the micro-structured fibres. The waveguide losses have been calculated through the imaginary parts of the propagation constant of the fundamental core mode by the above two methods for different geometrical parameters of the fibre and the wavelengths. Further, using the calculated values of the core mode fields, we computed the distribution of the transverse component of the Poynting vector and its stream lines. The positions of the optical vortices in the cross – section of the micro – structured fibre were determined from the equation for the streamlines of the transverse component of the Poynting vector of the core mode $x/P_x = y/P_y$. The conditions $P_x(x,y) = P_y(x,y) = 0$ defined the singular points and lines of the streamline pattern in the cross section of the fibre. Optical vortices have been detected in the regions and individual points, where the lines of zero values of transverse Poynting vector components overlap or intersect. The analysis of the vortex distributions and loss dependencies allows us to identify us the connection between the structure of the vortices and the minimum loss values in the micro-structured fibres.